\pdfoutput=1
\documentclass[11pt]{arxiv-article}
	
\bibliography{References}

\usepackage{Definitions}

\usepackage{pifont}
\usepackage{hyperref}
\usepackage{mathrsfs}  
\newcommand*{\threeloop}{\includegraphics[width=.5cm]{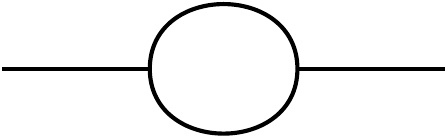}}
\newcommand*{\fourloop}{\includegraphics[width=.5cm]{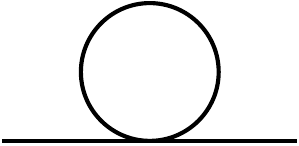}}

\def\mathclap#1{\text{\hbox to 0pt{\hss$\mathsurround=0pt#1$\hss}}}

\allowdisplaybreaks

\newcommand{\OfficialTitle}{
  The large-charge expansion for Schrödinger systems
}
\title{\setstretch{1.4}
  {\color{Thoughtless}\Huge\textbf{\dosserif\OfficialTitle}}
}

\hypersetup{pdfauthor={Samuel Favrod and Domenico Orlando and Susanne Reffert},pdftitle={\OfficialTitle}}

\author{%
  \begin{minipage}{.97\linewidth}
    \vspace{1cm}
    \begin{center} \dosserif%
      {\small
         \textbf{Samuel~Favrod}\textsuperscript{\ding{71}},
         \textbf{Domenico~Orlando}\textsuperscript{\ding{72}\ding{73}} and
         \textbf{Susanne~Reffert}\textsuperscript{\ding{73}} 
         }
    \end{center}
    \vspace{1cm}
    \authorBlock{\ding{71}}{\dosserif{} Institut für Theoretische Physik, ETH Zurich,\\
      CH-8093 Zürich, Switzerland}
    \authorBlock{\ding{72}}{\dosserif{} INFN sezione di Torino | Arnold--Regge Center\\
      via Pietro Giuria 1, 10125 Turin, Italy}
    \authorBlock{\ding{73}}{\dosserif{} Albert Einstein Center for Fundamental Physics\\
      Institute for Theoretical Physics, University of Bern,\\
      Sidlerstrasse 5, CH-3012 Bern, Switzerland}
  \end{minipage}
}

\date{}

\begin{document}

\setstretch{1.2}

\numberwithin{equation}{section}

\begin{titlepage}

  \newgeometry{top=23.1mm,bottom=46.1mm,left=34.6mm,right=34.6mm}

  \maketitle

  \thispagestyle{empty}

  \vfill\dosserif{}

  \abstract{\normalfont{}\noindent{}%
    In this note, we perform the large-charge expansion for non-relativistic systems with a global $U(1)$ symmetry in $3+1$ and $2+1$ space-time dimensions, motivated by applications to the unitary Fermi gas and anyons. These systems do not have full conformal invariance, but are invariant under the Schrödinger group. Also here, the low-energy physics is encoded by a Goldstone boson which is due to the breaking of the global symmetry when fixing the charge. %
    We find that in $2+1$ dimensions and higher, there is a large-charge expansion in which quantum corrections are suppressed with respect to the next-to-leading order terms in the Lagrangian. We give the next-to-leading-order expressions for the ground state energy and the speed of sound.
  }

\vfill

\end{titlepage}

\restoregeometry{}

\setstretch{1.2}

\section{Introduction}%
\label{sec:introduction}

The large-charge approach consists in studying \acp{cft} in sectors of fixed and large global charge, allowing a perturbative expansion of a generically strongly coupled theory with the inverse charge acting as a controlling parameter. It has been successfully applied to a variety of \acp{cft} in the last few years, such as $O(N)$ vector models~\cite{Hellerman:2015nra, Alvarez-Gaume:2016vff, Loukas:2016ckj, Monin:2016jmo, Hellerman:2017efx, Hellerman:2018sjf}, matrix models~\cite{Loukas:2017lof, Loukas:2017hic} and \acp{scft}~\cite{Hellerman:2017sur, Hellerman:2017veg, Hellerman:2018xpi, Bourget:2018obm}. Other aspects of \acp{cft} at large charge have been explored in~\cite{Cuomo:2017vzg, Jafferis:2017zna, Loukas:2018zjh}.
The predictions of the large-charge expansion have been independently verified by numerical methods to very high accuracy~\cite{Banerjee:2017fcx, delaFuente:2018qwv}. 

All the systems studied so far at large charge are relativistic. In this note, we apply the large-charge approach to non-relativistic systems. The most famous example of this class of systems is the four-dimensional unitary Fermi gas (for a review, see~\cite{randeria2012bcs}), which derives its importance from the fact that it can be experimentally realized with cold atoms~\cite{refId0}.
The unitary Fermi gas can be described by a non-relativistic superfluid and an effective Lagrangian was first proposed in~\cite{Son:2005rv}. In consequence, its low-energy dynamics is encoded by a Goldstone boson, just like in the bosonic theories studied previously at large charge.
Also anyons in three dimensions are described in the \ac{uv} by a Schrödinger particle minimally coupled to a Chern--Simons $U(1)$ theory~\cite{Chen:1989xs, Jackiw:1992fg, Bergman:1993kq, Nishida:2007pj}. Unlike in the treatment of \acp{cft} at large charge which are analyzed on $\mathbb{R} \times S^d$ in order to invoke the state-operator correspondence, we work throughout in flat space.\footnote{A state-operator correspondence also exists for non-relativistic \acp{cft}~\cite{Nishida:2007pj}; it necessitates studying the system in a harmonic potential. Non-relativistic \acp{cft} in a harmonic potential have been discussed at large charge in~\cite{Kravec:2018qnu} after this work was completed.} 

\medskip

Non-relativistic systems do not have the full conformal symmetry, but instead obey Schrödinger symmetry~\cite{Nishida:2007pj}. The main difference to the fully conformal case is the appearance of a mass scale $m$, which starkly modifies the type of terms that can appear in the Wilsonian effective action.

The leading-order piece of the effective action can be found via thermodynamic arguments~\cite{Greiter:1989qb}. We follow however (like in the original $O(2)$-model case) a reasoning based on dimensional analysis, which yields the same result in a simple and transparent fashion. In the non-relativistic case, some care must however be taken as natural units ($c=1, \ \hbar=1$) are not adapted to this limit. Since here, $c=\infty$, we will keep track of $\hbar$ all the way.

\medskip
We start out with the massive Schrödinger particle with a potential term which is compatible with Schrödinger symmetry. Since the Schrödinger field is a complex scalar, it can be written as $\psi= a e^{i\theta}$, making the global $U(1)$ symmetry $\theta \to \theta - \alpha$ manifest. We then identify the classical ground state at fixed and large charge and expand around it. Fixing the $U(1)$ charge $Q$ breaks the global $U(1)$ symmetry and the fluctuation $\chi$ around the minimum $\theta=\mu t$ acts as a Goldstone field. Once we have integrated out the non-propagating radial mode $a$, we find the same leading effective Lagrangian which was identified in~\cite{Son:2005rv} for the unitary Fermi gas. %

\medskip
Motivated by these findings, we set out to directly construct an effective Lagrangian for the Goldstone field, expanded around the ground state $\theta=\mu t$.
Our strategy for writing down the effective Lagrangian at large charge for the Goldstone field consists in the following three steps:
\begin{itemize}
	\item Using dimensional analysis, we write down the allowed terms under the assumption that there are no dimensionful couplings (apart from $m$).
	\item We compute the charge-scalings of these terms and only retain the ones with positive charge scaling, reducing the effective Lagrangian to only a handful of terms which are not suppressed in the limit of large charge.
	\item We use Schrödinger invariance to further constrain the terms that can appear in the effective Lagrangian.
\end{itemize}
We find two subleading higher-derivative terms at tree level at \ac{nlo} in the expansion in $\mu$.
The resulting effective Lagrangian takes for $2+1$ and $3+1$ dimensions the form
\begin{equation}
\begin{aligned}
  \mathcal{L}(\theta) ={}& c_0 \hbar^{1-d/2} m^{d/2} U^{(d+2)/2} \\
  &+ c_1 \hbar^{2-d/2} m^{-1+d/2} U^{(d-4)/2} \del_i U \del_i U \\
  &+ c_2 \hbar^{3-d/2}m^{-2+d/2} U^{(d-2)/2} (\del_i\del_i\theta)^2 + \order{\mu^{-2}},%
\end{aligned}
\end{equation}
where
\begin{equation}
  U = {\del_t \theta - \frac{\hbar}{2m}\del_i\theta\del_i\theta} .
\end{equation}
This is to be understood as an expansion around the classical ground state $\theta = \mu t$ and the charge density $\rho$ is related to the chemical potential via
\begin{equation}
	\mu = k\frac{d+2}{d}\frac{\hbar}{m}\rho^{{2}/{d}}.
\end{equation}
This matches for $3+1$ dimensions the result of~\cite{Son:2005rv}. 

The first quantum correction to this semi-classical result is the Casimir energy, which is proportional to $Q^{1/d}$.
Since no other term displays the same scaling, we predict that this is the only possible contribution at this order.

\medskip
Apart from the practical motivation stemming from the unitary Fermi gas, the exercise of writing down a large-charge expansion for a non-relativistic system is worthwhile, as our intuition of what an effective theory of a Goldstone boson should look like is heavily based on relativistic ideas. We come across many features which at first sight seem surprising, but are ultimately due to non-relativistic nature of our model.\footnote{For a general treatment of non-relativistic \acl{eft}, see~\cite{Leutwyler:1993gf}.}
It also serves to sharpen our ideas of the exact limits of applicability of the large-charge approach.
Just as for relativistic theories, where the large-charge approach is limited to space-time dimensions bigger or equal than $2+1$ by the Coleman--Mermin--Wagner theorem~\cite{Mermin:1966fe,Coleman:1973ci}, we find in the non-relativistic case that the expansion breaks down for dimensions smaller or equal than $1+1$.

\bigskip
This note is organized as follows. In Sec.~\ref{sec:schroedinger}, we discuss the Schrödinger particle at large charge, first performing the semi-classical analysis in Sec.~\ref{sec:semi-class} in which we find the ground state at large charge, integrate out the radial field $a$ and find the leading-order effective Lagrangian for the phase $\theta$. In Sec.~\ref{sec:quantum}, quantize the fluctuations $\eta$, $\hat\chi$ canonically in order to show the suppression of interaction terms by the large charge.
In Section~\ref{sec:schroedinger-eff} we directly construct an effective Lagrangian for the Goldstone boson alone.
In Section~\ref{sec:quantum-corr} we estimate the size of the quantum (loop) corrections to the tree-level propagator for the Goldstone $\chi$ and show that they are suppressed by inverse powers of $\mu$.
In Section~\ref{sec:four} we discuss the higher-derivative corrections at tree level in four dimensions and in Section~\ref{sec:three} in three dimensions, finding in both cases one term at order $\mu^0$ and that all further corrections are suppressed by at least $\mu^{-1}$.
In Section~\ref{sec:conclusions}, we present our final results, namely the energy of the ground state on the torus and the \ac{nlo} expression for the speed of sound, and end with concluding remarks and an outlook.
In Appendix~\ref{sec:schroedinger-symm} we collect the necessary background material on Schrödinger symmetry.

\section{The Schrödinger particle at large charge}%
\label{sec:schroedinger}

\subsection{Setup}
While it is possible to directly construct the effective Lagrangian for a Goldstone boson at large charge, which we will do in the next section, we first construct the leading order starting from the simplest non-relativistic system, namely the Schrödinger Lagrangian in $d+1$ dimensions with a potential $V$, 
\begin{equation}\label{eq:Schroedinger_lag}
\mathcal{L}(\psi)=  \frac{i}{2} ( \psi^*\partial_t \psi - \psi \partial_t \psi^*) -\frac{\hbar}{2m}\partial_i \psi^* \partial_i \psi - V(\psi^* \psi),
\end{equation}
where we disregarded possible higher-derivative terms.
The form of the potential term can be determined entirely by dimensional analysis, under the assumption that there are no dimensionful parameters except $m$:
\begin{equation}\label{eq:extrapotSchroedinger}
V(\psi^*\psi) = \frac{k}{m} \hbar^{\frac{d-2}{d}} (\psi^*\psi)^{\frac{d+2}{d}},
\end{equation}
with $k > 0$ a dimensionless constant. $V$ is invariant under the 13-parameter Schrödinger group generated by the operators in Eq.~\eqref{eq:Schroedinger-generators} in the appendix.

The global $U(1)$ symmetry of the system, $\psi \rightarrow e^{-i\alpha} \psi$, has associated Noether charge density $\rho$ given by\footnote{The minus sign in the $U(1)$ transformation is necessary in order to obtain a positive charge.} 
\begin{equation}
\rho = \frac{1}{\hbar} \left(\frac{\partial \mathcal{L}}{\partial(\partial_t \psi)}\delta \psi+ \frac{\partial \mathcal{L}}{\partial(\partial_t \psi^*)}\delta \psi^*\right) = \frac{1}{\hbar}\psi^*\psi.
\end{equation}
We fix this charge via the condition
\begin{equation}
\int \dd[d]{x} \rho = Q,
\end{equation}
which we choose to fix such that $Q \gg 1$. 

We now look for a ground state at fixed charge which is homogeneous in space, so $\rho=Q/\text{volume}$ is constant.\footnote{Such a state, should it exist, is always of lower energy than a non-homogeneous state.}
We solve the \ac{eom}
\begin{equation}
i\partial_t \psi = \frac{\partial V}{\partial \psi^*}
\end{equation}
and find 
\begin{equation}
\psi = A e^{-i \mu t},
\end{equation}
 with $A = \sqrt{\hbar\rho}$ and 
\begin{equation}
	\mu = k\frac{d+2}{d}\frac{\hbar}{m}\rho^{{2}/{d}},
\end{equation} 
where $\mu$ plays the role of a chemical potential and due to the above scaling will be large for $Q$ large.

The energy density of the classical ground state is given by
\begin{equation}\label{eq:cl-gr-st-en}
  \mathcal{E}_{0} = \hbar^2 \frac{k}{m} \rho^{(d+2)/d} \propto \mu^{(d+2)/2}.
\end{equation}

\subsection{Semi-classical analysis}%
\label{sec:semi-class}

Choosing a fixed-charge ground state spontaneously breaks the global $U(1)$ symmetry, leading to the appearance of a massless mode, the Goldstone boson~\cite{Alvarez-Gaume:2016vff}. To see this, we now expand the Lagrangian~\eqref{eq:Schroedinger_lag} to second order in the fluctuations around the ground state we identified above, writing the field as
\begin{equation}
  \psi(t,x) = a(t,x) e^{-i\theta(t,x)} = \left(\sqrt{\hbar\rho}+\eta(t, x)\right)e^{-i(\mu t + \hat\chi(t,x))}, 
\end{equation}
where $\eta(t, x), \ \hat\chi(t,x)$ are the fluctuations around the ground state. The quadratic terms in the fields are given by
\begin{equation}\label{eq:L2}
  \mathcal{L}_2(\eta, \hat\chi) = \sqrt{\hbar \rho}\left(\eta \dot{\hat\chi} -\dot\eta {\hat\chi}\right) - \frac{\hbar}{2m}\del_i \eta \del_i \eta - \frac{\hbar^2\rho}{2m}\del_i\hat\chi\del_i\hat\chi -  \frac{4}{d}\mu \eta^2,
\end{equation}
where for later convenience we have added a surface term to obtain a time derivative for $\eta$.

The dispersion relation is found via the inverse propagator,
\begin{equation}
  \begin{vmatrix}
    \frac{\hbar}{2m} p^2 + \frac{4}{d} \mu & -i\sqrt{\hbar\rho}\,\omega \\
    i\sqrt{\hbar\rho}\,\omega & \frac{\hbar^2\rho}{2m}p^2
  \end{vmatrix} = 0,
\end{equation}
so
\begin{equation}
  \label{eq:dispersion-relation}
  \omega(p) = \sqrt{\frac{2}{d}}\sqrt{\frac{\hbar \mu}{m}}\abs{p} \left( 1 + \frac{d}{16}\frac{\hbar}{m\mu} p^2 \right) + \order{\mu^{-3/2}}.
\end{equation}
As expected, the dispersion relation shows linear behavior in $p$ near zero even though the system is non-relativistic. The speed of sound is given by
\begin{equation}\label{eq:speedsound}
	c_s^2 = {\frac{2}{d}}{\frac{\hbar \mu}{m}},
\end{equation}
differently from the relativistic system in which scale invariance fixes the speed of sound to $(c_s^{\text{rel}})^2 = c^2/d$~\cite{Hellerman:2015nra}.

\medskip
An important difference to the relativistic cases described so far in the literature, where the radial mode was massive, is that here, the radial mode does not propagate. We can therefore again integrate it out and write an effective Lagrangian only for the Goldstone boson $\hat\chi$, at least to leading order. To lowest order, we can use the saddle point approximation and express the radial mode $a$ via its \ac{eom} in terms of $\theta$. Assuming that $a$ varies slowly, we find
\begin{equation}\label{eq:radial}
  a(t,x)^{d/4} \approx \frac{d}{d+2}\frac{m \hbar^{(2-d)/d}}{k}\left( \del_t \theta - \frac{\hbar}{2m}\del_i\theta\del_i\theta \right).
\end{equation}
Substituting this back into the Lagrangian~\eqref{eq:Schroedinger_lag}, we find
\begin{equation}\label{eq:Ltheta-leading}
	\mathcal{L}^{(0)}({\theta}) \propto %
	\hbar^{(2-d)/2} m^{d/2}\left( \del_t \theta - \frac{\hbar}{2m}\del_i\theta\del_i\theta \right)^{(d+2)/2}.
\end{equation}

In Section~\ref{sec:schroedinger-eff}, we will construct the subleading corrections in the large charge. This reproduces the leading-order effective Lagrangian of the four-dimensional unitary Fermi gas, which can also be obtained via the thermodynamic reasoning of~\cite{Greiter:1989qb,Son:2005rv}.

\subsection{Quantization}%
\label{sec:quantum}

In the following, we want to quantize the fluctuations $\eta$, $\hat\chi$ canonically in order to show the suppression of interaction terms by the large charge.

We first need to perform a Legendre-transform of $\mathcal{L}_2$ in~\eqref{eq:L2}, using
\begin{align}
	\Pi_\eta &= \frac{\delta \mathcal{L}_2}{\delta \dot\eta}  = - \sqrt{\hbar\rho} \,\hat\chi, & \Pi_{\hat\chi} &= \frac{\delta \mathcal{L}_2}{\delta \dot{\hat\chi}}  = \sqrt{\hbar\rho}\, \eta,
\end{align}
which results in
\begin{equation}\label{eq:H2}
	\mathcal{H}_2 = \frac{\delta \mathcal{L}_2}{\delta \dot\eta} \dot\eta + \frac{\delta \mathcal{L}_2}{\delta \dot{\hat\chi}}\dot{\hat\chi} - \mathcal{L}_2 =
	 \frac{\hbar}{2m}\del_i \eta \del_i \eta + \frac{\hbar^2\rho}{2m}\del_i\hat\chi\del_i\hat\chi +  \frac{4}{d}\mu \eta^2 .
\end{equation}
The \ac{eom} associated to $\mathcal{H}_{(2)}$ are
\begin{equation}
  \label{eq:schrodinger_largecharge_2}
  \begin{cases}
    \dot {\hat\chi} = - \frac{\hbar^{1/2}}{2m \rho^{1/2}} \nabla^2 \eta + \frac{4 \mu }{d \sqrt{\hbar \rho}} \eta , \\
    \dot \eta = \frac{\hbar^{3/2} \rho^{1/2}}{2m} \nabla^2 \hat\chi .
  \end{cases}
\end{equation}
To canonically quantize the Hamiltonian, we impose the usual equal-time commutation relations,
\begin{equation}
  \comm{\phi(t,x)}{\Pi_\phi(t,y)} = i \hbar \,\delta(x-y),
\end{equation}
which in our case take the form
\begin{equation}
  \comm{\hat\chi(t,x)}{\eta(t,y)} = i \sqrt{\frac{\hbar}{\rho}}\,\delta(x-y).
\end{equation}

In order to rewrite everything in terms of standard creation and annihilation operators, it is convenient to make a choice of what we call field and what we call momentum.
For example, using \(\hat \chi\) and \(\Pi_{\hat \chi}\), the Hamiltonian density becomes
\begin{equation}
  \mathcal{H}_2 = \frac{1}{2 m \rho} \del_i \Pi_{\hat \chi} \del_i \Pi_{\hat \chi} + \frac{\hbar^2 \rho}{2 m} \del_i \hat \chi \del_i \hat \chi + \frac{4}{\hbar \rho d} \Pi_{\hat \chi}^2
\end{equation}
Then we can write
\begin{align}
  \hat \chi(x) &= \int \frac{\dd[d]p}{(2 \pi)^d} \frac{1}{\sqrt{2 \lambda}} \pqty{a_p + a_{-p}^\dagger } e^{i p x} , \\
  \Pi_{\hat \chi}(x) &= \int \frac{\dd[d]p}{(2 \pi)^d} (-i) \sqrt{\frac{\lambda}{2}} \pqty{a_p - a_{-p}^\dagger} e^{i p x}
\end{align}
where \(\lambda \) is a function of \(p\), and the \(a_p\) are standard creation/annihilation operators.
\begin{equation}
  \comm{a_p}{a_{q}^{\dagger}} = (2\pi)^d \delta(p-q).
\end{equation}
These expressions automatically satisfy the commutation relations of \(\hat \chi\) with its momentum.
We only need the value of \(\lambda(p)\) that diagonalizes the Hamiltonian.
After a standard calculation we find that if
\begin{equation}
  \lambda(p) =\frac{\hbar \rho p^2}{2 m \omega(p)} ,
\end{equation}
where \(\omega(p)\) is the expression in Eq.~(\ref{eq:dispersion-relation}), the Hamiltonian becomes
\begin{equation}
  \mathcal{H}_2  = 2\int \frac{\dd[d]{p}}{(2\pi)^{d}}   \omega(p)\left( a_p^{\dagger}a_p + \tfrac{1}{2}\comm{a_p}{a^\dagger_p}\right) .
\end{equation}

What happens to the interaction terms in this picture?
Using the form of \(\Pi_{\hat \chi}\), we can see that the field \(\eta\) scales as \(\eta \sim \mu^{-1/4}\).
Then, expanding the potential term \(V(\eta^2)\) in the action, we see that the generic term in the expansion in \(\mu\) scales as
\begin{equation}
  V(\eta) = \sum_{n=2}^\infty c_n \frac{1}{\mu^{(n d - d - 4)/4}} \eta^n = \sum_{n=2}^\infty c_n \frac{1}{\mu^{(n (d + 1 ) - d - 3)/4}} (\eta')^n,
\end{equation}
where \(\eta'\) is rescaled to be of order \(\mu^0 = 1\).
This shows explicitly that the interactions are parametrically suppressed with respect to the quadratic part of the Hamiltonian which, in this normalization is of order \(\mu^{1/2}\).

\section{The effective action for the Goldstone field}%
\label{sec:schroedinger-eff}

\subsection{Leading order}
The Goldstone field that we have identified in the preceding section dominates the low-energy physics of our non-relativistic model at large charge. It is therefore convenient to directly construct an effective action for this field alone. 

The leading part of the effective Lagrangian~\eqref{eq:Ltheta-leading} can also be constructed using invariance under Schrödinger symmetry:
\begin{equation}\label{eq:Ltheta-leading-2}
	\mathcal{L}^{(0)}(\theta) = c_0 %
	\hbar^{(2-d)/2} m^{d/2}\left( \del_t \theta - \frac{\hbar}{2m}\del_i\theta\del_i\theta \right)^{(d+2)/2}= c_0 %
	\hbar^{(2-d)/2} m^{d/2}U^{(d+2)/2},
\end{equation}
where $c_0$ is a dimensionless constant. 
The \ac{eom} at leading order is given by
\begin{equation}\label{eq:eom-LO}
\del_t U - \frac{\hbar}{m}\del_i U \del_i \theta - \frac{2\hbar}{md}U \del_i\del_i \theta = 0.	
\end{equation}
The $U(1)$ charge density is
\begin{equation}
	\rho = \frac{\delta \mathcal{L}_{\theta}^{(0)}}{\delta\dot\theta} = \frac{d+2}{2}c_0 m^{d/2} \hbar^{(2-d)/2}U^{d/2},
\end{equation}
and the solution to the classical \ac{eom} is $\theta = \mu t$, where $\mu$ is given by
\begin{equation}
	\rho = \frac{d+2}{2}c_0 m^{d/2} \hbar^{(2-d)/2}\mu^{d/2}.
\end{equation}
We see that we can equivalently use $\mu$ as an expansion parameter as $\mu$ is given by a positive power of $\rho$.
Let us expand $\theta$ around the ground state $\theta = \mu t + \kappa \chi$, where $\chi$ is the Goldstone field and its normalization is chosen such that the kinetic term $\del_t \chi\del_t \chi$ is canonical:
\begin{equation}\label{eq:kappa}
    \kappa = 2 d_0 m^{-d/4}\hbar^{(d-2)/4}\frac{1}{\mu^{(d-2)/4}}
    = \frac{2^{(d+1)/d}}{(d+2)^{1/d}c_0^{1/d}}\frac{\hbar^{(d-2)/(2d)}}{\sqrt{dm}}\frac{1}{\rho^{(d-2)/(2d)}}.
\end{equation}
where we have introduced \(d_0^2 = 2/(c_0 d(d+2))\).

We expect that the cases $d \gtrless 2$ will lead to very different outcomes. For $d>2$, the fluctuations are parametrically small in the large-charge expansion, allowing us to perform perturbative calculations. In the following, we will however see that also the case $d=2$ is well-behaved.

We expand the Lagrangian~\eqref{eq:Ltheta-leading-2} (up to a boundary term) around the ground state solution:
\begin{equation}\label{eq:Lchi-leading}
  \begin{aligned}
    \mathcal{L}^{(0)}({\chi}) &= c_0 \hbar^{(d-2)/2} m^{d/2} \mu^{(d+2)/2} + \del_t\chi\del_t\chi - \frac{2}{d}\frac{\hbar \mu}{m} \del_i\chi \del_i\chi + \order{\frac{1}{\mu^{(d-2)/4}}}\\
     & = \left(\textstyle{\frac{2}{d+2}}\right)^{(d+2)/d} \frac{\hbar^{(d-2)/d}}{m c_0^{d/2}}\rho^{(d+2)/d} + \del_t\chi\del_t\chi - c_s^2 \del_i\chi \del_i\chi + \order{\frac{1}{\rho^{(d-2)/(2d)}}},
  \end{aligned}
\end{equation}
where we have recovered the speed of sound of Eq.~\eqref{eq:speedsound}.

There will be two kinds of corrections to the above leading-order semi-classical effective Lagrangian: quantum loop corrections and higher-derivative terms at tree-level.
As the behavior of the system is different for $d=3$ and $d=2$, we will discuss the higher-derivative terms of these cases separately.
For \(d<2\), we have seen that the fluctuations grow with the charge.
This means that the higher-derivative terms are \emph{not suppressed} but enhanced for large charge, thus breaking our construction.

\subsection{Quantum corrections}%
\label{sec:quantum-corr}

In the following, we want to estimate the size of the quantum (loop) corrections to the tree-level propagator for the Goldstone $\chi$. We want to show in particular that they are suppressed by inverse powers of $\mu$. In order to do so, we need to continue the expansion of the effective Lagrangian to include interaction terms of up to order four:
\begin{align}  \label{eq:Lchi-leading-int}
    \mathcal{L}^{(0)}({\chi}) ={}& \left(\frac{2}{d+2}\right)^{(d+2)/d} \frac{\hbar^{(d-2)/d}}{m c_0^{d/2}}\rho^{(d+2)/d} + \del_t\chi\del_t\chi - c_s^2 \del_i\chi \del_i\chi \\ \nonumber
    & -2 \pqty{\frac{2}{c_0 d(d+2)}}^{1/2} \frac{\hbar^{(d+2)/4}}{m^{(d+4)/4}} \frac{1}{\mu^{(d -2)/4}} \del_t \chi \del_i \chi \del_i \chi \\ \nonumber
    &+  \frac{d-2}{3} \sqrt{\frac{2}{c_0 d(d+2)}} \frac{\hbar^{(d-2)/4}}{{m^{d/4}}} \frac{1}{\mu^{( d+ 2)/4}} (\del_t \chi)^3 \\ \nonumber
    & - \frac{2(d-2)}{c_0 d(d+2)} \frac{\hbar^{d/2}}{ m^{(d+2)/2}}\frac{1}{\mu^{d/2}} (\del_t \chi)^2 \del_i \chi \del_i \chi \\ \nonumber
    & + \frac{2}{c_0d(d+2)} \frac{\hbar^{(d+2)/2}}{m^{(d+2)/2}} \frac{1}{\mu^{(d - 2)/2}} \del_i \chi \del_i \chi \del_j \chi \del_j \chi + \dots. \nonumber
\end{align}
Since the speed of sound depends explicitly on the chemical potential \(\mu\), in this section it is convenient to rescale the space coordinates so that \(c_s \del_i = \del_i'\).
The action then takes the form
\begin{equation}\label{eq:Lag-0-chi}
  \begin{aligned}
    \mathcal{L}^{(0)}(\chi) ={}& \left(\frac{2}{d+2}\right)^{(d+2)/d} \frac{\hbar^{(d-2)/d}}{m c_0^{d/2}}\rho^{(d+2)/d} + \del_t \chi \del_t \chi - \del_i' \chi \del_i' \chi \\
    &+ d_0 \frac{\hbar^{(d-2)/4}}{m^{d/4} \mu^{(d+2)/4}} \pqty{ \frac{d-2}{3} (\del_t \chi)^3 - d \del_t \chi \del_i' \chi \del_i' \chi } \\
    &+ \frac{d_0^2}{4} d \frac{\hbar^{(d-2)/2}}{m^{d/2} \mu^{(d+2)/2}} \pqty{ d \del_i' \chi \del_i' \chi \del_j' \chi \del_j' \chi - 2(d-2) (\del_t \chi)^2 \del_i' \chi \del_i' \chi }.
\end{aligned}
\end{equation}
The Feynman rules are easily computed starting from this effective Lagrangian.
\begin{itemize}
\item The tree-level propagator is given by
\begin{equation}
	D(\omega,p) = \frac{1}{\omega^2 - p^2}.
\end{equation}
\item The trivalent vertex has the form
  \begin{equation}
    \propto d_0 \frac{\hbar^{(d-2)/4}}{m^{d/4} \mu^{(d+2)/4}} \pqty{\frac{d-2}{3} \omega_1 \omega_2 \omega_3 - d \omega_1 p_2 \cdot p_3 } + \text{permutations}.
\end{equation}
\item The quadrivalent vertex has the form
  \begin{equation}
    \propto \frac{d_0^2}{4} d \frac{\hbar^{(d-2)/2}}{m^{d/2} \mu^{(d+2)/2}} \pqty{ d(p_1 \cdot p_2)(p_3 \cdot p_4) - 2(d-2) \omega_1 \omega_2 p_3 \cdot p_4} + \text{permutations}.
  \end{equation}
\end{itemize}
From these building blocks we can construct the following two one-loop diagrams:
\begin{align}
  \includegraphics{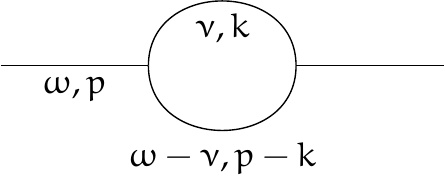} &&  \includegraphics{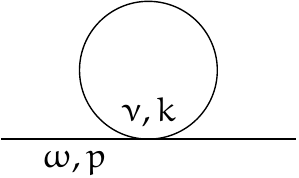} 
\end{align}
We now want to evaluate how the coupling \(g = g_0 + g_1 + \dots\) of the quadratic part \((\del_t \chi)^2 - \del_i' \chi \del_i' \chi\) in the Lagrangian is renormalized at one loop, where  $g_0  = 1  $ in Eq.~\eqref{eq:Lag-0-chi} and $g_1$ is the one-loop contribution.

We compute the self-energy \(\Pi(\omega, p)\) associated to the two diagrams above and use the \ac{rg} flow equations to compare the ratio of the one-loop contribution to the tree-level term,
\begin{equation}
  \frac{\Lambda \fdv{g_1 }{\Lambda}}{g_0} = D(\omega,p) \Lambda \fdv{\Pi(\omega,p)}{\Lambda}.
\end{equation}

Since we want to estimate the \(\mu\)-scaling of these effects it is convenient to separate the dimensionful part of the self-energy from the dimensionless one:
\begin{equation}
  \Pi(\omega, p) = \mu^{\alpha_\mu} \hbar^{\alpha_\hbar} m^{\alpha_m} \Pi'(\omega, p),
\end{equation}
where the \(\alpha_i\) are exponents to be computed and \(\Pi'\) is the self-energy in which we set \(\mu = \hbar = m = 1\).

In the case of the two one-loop diagrams it is easy to see that both self-energies scale in the same way:
\begin{equation}
  \begin{aligned}
    \Pi_{\threeloop}(\omega, p ) &= \frac{\hbar^{(d-2)/2}}{m^{d/2} \mu^{(d+2)/2}} \Pi'_{\threeloop}(\omega, p ), \\
    \Pi_{\fourloop}(\omega, p ) &= \frac{\hbar^{(d-2)/2}}{m^{d/2} \mu^{(d+2)/2}} \Pi'_{\fourloop}(\omega, p ),
  \end{aligned}
\end{equation}
which means that at one loop we have 
\begin{multline}
    \frac{\Lambda \fdv{g_1 }{\Lambda}}{g_0} = D(\omega,p) \Lambda \fdv{\Pi(\omega,p)}{\Lambda} = D(\omega,p) \Lambda \fdv{\Lambda} \pqty{  \Pi_{\threeloop}(\omega, p ) +  \Pi_{\fourloop}(\omega, p ) } \\
    = \frac{\hbar^{(d-2)/2}}{m^{d/2} \mu^{(d+2)/2}}   D(\omega,p) \Lambda \fdv{\Lambda} \pqty{  \Pi'_{\threeloop}(\omega, p ) +  \Pi'_{\fourloop}(\omega, p ) } = \order{\frac{1}{\mu^{(d+2)/2}}} .
\end{multline}
We see that the one-loop effects are suppressed by negative powers of $\mu$ for all dimensions and that the suppression grows with higher space dimensions.

Now that we have the quantum corrections under control, we can move on to the corrections due to higher-derivative terms at tree level.

\subsection{Higher-derivative terms in four dimensions}%
\label{sec:four}

We first start with the obviously well-behaved case of $3+1$ spacetime dimensions. In the first step, we write down the allowed terms using dimensional analysis under the assumption that there are no dimensionful couplings other than $m$.

We have the following ingredients to generate higher terms:
\begin{equation}\label{eq:ingredients}
	m,\ \hbar,\ \del_t,\ \del_i, \theta, \ U=\del_t \theta - \frac{\hbar}{2m}\del_i\theta\del_i\theta,
\end{equation}
where the combination $U$ is suggested by the result of integrating out the radial mode~\eqref{eq:radial}. The most generic term with the right dimensions has the schematic form
\begin{equation}\label{eq:O}
	\mathcal{O}_{\alpha,\beta} \propto \hbar^{\beta-1/2} m^{3/2-\beta} \del_t^\alpha \del_i^{2\beta} U^{5/2-\alpha-\beta},
\end{equation}
where $\alpha$, $\beta$ are positive integers (by locality) and the derivatives can act either on $U$ or on the field $\theta$, which is dimensionless as it is a Goldstone field, and can thus appear only through its derivatives. 

We will also exclude terms which are odd under parity 
\begin{equation}
  (t, \theta) \to (-t, -\theta) .
\end{equation}

We can further constrain the form of $\mathcal{O}_{\alpha,\beta}$ by imposing invariance under the Schrödinger group, which means in particular that it has to be a Galilei scalar.
The key observation is that the Jacobian $\dd{t}\dd[3]{x}$ transforms homogeneously both under dilatation and \ac{sct} (for details see Appendix~\ref{sec:schroedinger-symm}):
\begin{align}
  \text{scale}&: \dd{t}\dd[d]{x} \to e^{(d+2) \tau} \dd{t}\dd[d]{x}, \\
  \text{\ac{sct}}&: \dd{t}\dd[d]{x} \to (1+\lambda t)^{-(d+2)} \dd{t}\dd[d]{x}.
\end{align}
This fixes the transformation properties of $\mathcal{O}_{\alpha,\beta}$. As shown in Appendix~\ref{sec:schroedinger-symm}, dimensional analysis implies scale invariance, so we can confine ourselves to verifying invariance under \ac{sct}.

The only covariant way in which the time derivative can appear is in the combination appearing in $U$. 
This means that we can set $\alpha=0$ in~\eqref{eq:O} and write the generic term as
\begin{equation}\label{eq:Ofin}
  \mathcal{O}_{\beta} \propto \hbar^{\beta-1/2} m^{3/2-\beta} \del_i^{2\beta} U^{5/2-\beta}.
\end{equation}
Since no time-derivatives appear explicitly, only terms with an even number of $\theta$ fields are allowed by parity.

At this point, it is convenient to use the $\mu$ scaling of the possible terms to determine those higher-derivative terms which are not suppressed in the limit of large charge.
We have the following leading $\mu$ scalings:
\begin{align}
  U &\sim \mu, & \del_i \theta &\sim \mu^{-1/4}, & \del_i U &\sim \mu^{-1/4}.
\end{align}
For  given $\beta$, the term with the highest $\mu$ scaling is obtained when all the space derivatives act on exactly two fields $\theta$:
\begin{equation}\label{eq:Omax}
  \mathcal{O}_{\beta}^{\text{max}} \propto \hbar^{\beta-1/2} m^{3/2-\beta} U^{5/2-\beta} \del_i^n \theta \del_i^m \theta\sim \mu^{2 - \beta},
\end{equation}
with $n+m=2\beta$.
We see that only for $\beta \leq 2$, the $\mu$ scaling can be non-negative.
In order to compare with the quantum corrections discussed in Section~\ref{sec:quantum-corr} it is convenient to rescale the space derivatives as \(\del'_i = \mu^{-1/2} \del_i\) so that the leading quadratic term is of order \(\mu^0\) as in Eq.~\eqref{eq:Lag-0-chi}. Using the rescaled space-derivates, the maximum scaling of the operators becomes $\mu^{2-2\beta}$.

We now check the possible terms explicitly:
\begin{itemize}
\item $\beta=0$ admits a unique term, namely the leading-order contribution 
  \begin{equation}
    \mathcal{O}_0 = \hbar^{-1/2} m^{3/2} U^{5/2}.
  \end{equation}
\item For $\beta = 1$, there are (up to total derivatives) two possible terms,
  \begin{align}
    \mathcal{O}_1^{(1)} &= \hbar^{1/2}m^{1/2} U^{-1/2} \del'_i U \del'_i U,\label{eq:O11-d3}\\
    \mathcal{O}_1^{(2)} &= \hbar^{1/2}m^{1/2} U^{3/2} \del'_i \theta \del'_i \theta.
  \end{align} 
  The second term is however excluded by \ac{sct}. We are thus left with only $\mathcal{O}_1^{(1)}$, which appears also in the effective Lagrangian of~\cite{Son:2005rv} which is an expansion in small momenta. This term however has $\mu$ scaling $\mathcal{O}_1^{(1)} \sim \mu^{-2}$ and is thus suppressed at large charge, but gives the first \ac{nlo} correction.
\item For $\beta = 2$, the only possible terms that naively admit non-negative $\mu$ scaling are
\begin{align}
  \mathcal{O}_2^{(1)} &= \hbar^{3/2}m^{-1/2} U^{1/2} \del'_i \del'_i \theta \del'_j \del'_j \theta,\label{eq:O21-d3}\\
  \mathcal{O}_2^{(2)} &= \hbar^{3/2}m^{-1/2} U^{1/2} \del'_i \theta \del'_i  \del'_j \del'_j \theta.
\end{align}
The operator $\mathcal{O}_2^{(1)}$ is only invariant on-shell, \emph{i.e.} its variation is proportional to the \ac{eom} at leading order~\eqref{eq:eom-LO} (after integrating by parts and neglecting a boundary term). It appears also in the effective Lagrangian of the unitary Fermi gas~\cite{Son:2005rv}. We see that $\mathcal{O}_2^{(1)}$ scales as $\mathcal{O}_2^{(1)} \sim \mu^{-2}$, and it  appears in the effective Lagrangian as a \ac{nlo} correction together with the term above.
$\mathcal{O}_2^{(2)}$ is not invariant under \ac{sct}, either on or off-shell.%
\item For $\beta >2$, the first term compatible with the symmetries is
\begin{equation}
	\mathcal{O}_3^{(1)} = \hbar^{5/2}m^{-3/2} U^{-1/2} \del'_i  \del'_i  \del'_j \theta \del'_j \del'_k  \del'_k \theta,
\end{equation}
with a (rescaled) $\mu$ scaling of $\mathcal{O}_3^{(1)} \sim \mu^{-4}$ and it is more suppressed than the two terms above.
\end{itemize}
We see that the first correction to the leading-order effective Lagrangian comes at order $\mu^{-2}$ and all further corrections are already suppressed by $\mu^{-4}$.

\subsection{Higher-derivative terms in three dimensions}
\label{sec:three}

Next, we will study higher-derivative corrections to the leading-order effective Lagrangian in $2+1$ space-time dimensions. We have as before the ingredients~\eqref{eq:ingredients} and the most generic term allowed by dimensional analysis now has the form
\begin{equation}\label{eq:O3}
  \mathcal{O}_{\alpha,\beta} \propto \hbar^{\beta} m^{1-\beta} \del_t^\alpha \del_i^{2\beta} U^{2-\alpha-\beta}.
\end{equation}
As in the four-dimensional case, time-derivatives of $\theta$ that do not appear inside \(U\) are excluded by Schrödinger symmetry, so we can set $\alpha=0$:
\begin{equation}\label{eq:O3red}
  \mathcal{O}_{\beta} \propto \hbar^{\beta} m^{1-\beta} \del_i^{2\beta} U^{2-\beta}.
\end{equation}
As before, we only allow parity-invariant terms.
We use again the $\mu$ scaling of the possible terms to determine those higher-derivative terms which are not suppressed in the limit of large charge.
We now have the following leading $\mu$ scalings:
\begin{align}
  U &\sim \mu, & \del_i \theta &\sim 1, & \del_i U &\sim 1.
\end{align}
Introducing again \(\del'_i = \mu^{-1/2} \del_i\) we see that for given $\beta$, the term with the highest $\mu$ scaling is
\begin{equation}
  \mathcal{O}_{\beta}^{\text{max}} \sim  \mu^{2 - 2 \beta} ,
\end{equation}
independently of where the derivatives act.
We see that again only for $\beta \leq 2$, the $\mu$ scaling can be non-negative. We check the possible terms explicitly:
\begin{itemize}
\item $\beta=0$ admits a unique term, namely the leading-order contribution 
  \begin{equation}
    \mathcal{O}_0 = m U^2.
  \end{equation}
  This term has \(\mu\)-scaling \(\mathcal{O}_0 \sim \mu^2\).
\item For $\beta = 1$, there is the term
  \begin{align}
    \mathcal{O}_1 &= \hbar U^{-1} \del'_i U \del'_i U.
  \end{align}
  This term has \(\mu\)-scaling \(\mathcal{O}_1 \sim \mu^{-2}\).
\item For $\beta = 2$, the only possible term that naively admits non-negative $\mu$ scaling is
  \begin{align}\label{eq:O21-d2}
    \mathcal{O}_2 &= \hbar^{2}m^{-1} \del'_i \del'_i \theta \del'_j \del'_j \theta.
  \end{align}
  This term is invariant under the Schrödinger group (up to a total derivative) and has again scaling $\mathcal{O}_2 \sim \mu^{-2}$, thus contributing to the effective Lagrangian at the same order as \(\mathcal{O}_1\).
\end{itemize}
We find thus, that also in three dimensions, the large-charge expansion works and the \ac{nlo} correction terms have scaling $\mu^{-2}$.
This three-dimensional discussion may apply to the case of anyons, where in the \ac{uv}, the system is described by a Schrödinger particle minimally coupled to a Chern--Simons $U(1)$ theory~\cite{Chen:1989xs, Jackiw:1992fg, Bergman:1993kq, Nishida:2007pj}.

\section{Results and Conclusions}%
\label{sec:conclusions}

In this note, we have studied the large-charge expansion of Schrödinger-invariant non-relativistic systems with a global $U(1)$ symmetry in $2+1$ and $3+1$ space-time dimensions. As expected, the fixed-charge low-energy dynamics is governed by a Goldstone boson. We found in both cases that apart from two \ac{nlo} terms, all further corrections both from higher-derivative terms at tree-level and quantum corrections are suppressed by higher inverse powers of $\mu$. The resulting effective Lagrangian takes the form
\begin{equation}
\begin{aligned}
  \mathcal{L}(\theta) ={}& c_0 \hbar^{1-d/2} m^{d/2} U^{(d+2)/2} \\
  &+ c_1 \hbar^{2-d/2} m^{-1+d/2} U^{(d-4)/2} \del_i U \del_i U \\
  &+ c_2 \hbar^{3-d/2}m^{-2+d/2} U^{(d-2)/2} (\del_i\del_i\theta)^2 + \order{\mu^{-2}},%
\end{aligned}
\end{equation}
where the $c_i$ are dimensionless Wilsonian couplings.
The resulting dispersion relation is given by
\begin{equation}
  \omega = c_s p \pqty{ 1 - d_0^2 \frac{\hbar}{m} \pqty{2 c_1 + d c_2} \frac{p^2}{\mu} + \order{\frac{1}{\mu^2}}},
\end{equation}
which again matches the result of~\cite{Son:2005rv}. We find an \(\order{\rho^0}\) correction to the speed of sound \(c_s^2 = 2 \hbar \mu/ (m d) \), while quantum corrections enter only at a higher order.

Based on the above effective Lagrangian, we can now give as an example of an observable the expression for the energy of the ground state on the torus,
\begin{equation}
  E_{T^d} = \frac{\hbar^2}{m} \bqty{ V b_1^2 \rho^{(d+2)/d} + \frac{b_1}{V^{1/d} d} \sqrt{\frac{d+2}{2}} \rho^{1/d}\zeta_{T^d}(-2) + \frac{b_2}{V^{2/d}}} + \order{\frac{1}{\rho^{2/d}}},
\end{equation}
where $b_1$ and \(b_2\) are constants related to the $c_i$ above, $V$ is the volume of the torus $T^d$, and $\zeta_{T^d}$ is the Zeta-function of a unit square torus.
We see that the energy receives a contribution from the classical ground state energy~\eqref{eq:cl-gr-st-en} and from the Casimir energy.
Note that the four-derivative terms $\mathcal{O}_1$ and \(\mathcal{O}_2\) do not contribute to the energy of a homogeneous configuration as they contain space derivatives, but enter the Casimir energy at order \(\order{\rho^0}\) with a coefficient \(b_2\) that can only be computed numerically.
All other classical and quantum corrections are suppressed by inverse powers of $\rho$.\footnote{In particular the first quantum correction enters at order \(\order{\mu^{-(d+4)/2}}\).}

\medskip

The main difference between the model studied in this note and the relativistic systems that have been discussed so far in the limit of large charge in the literature is the presence of a dimensionful parameter \(m\), which we can identify \emph{e.g.} with the Fermi mass in the case of fermions at unitarity.
This parameter is compatible with the Schrödinger symmetry (where it can be thought of as a central charge) and we do not need to make any assumptions about it, \emph{i.e.} it is \emph{not} used to write a derivative expansion.
The terms that are allowed by the symmetry are quite different with respect to the relativistic case and lead to a characteristic signature in the dependence of the energy on the \(U(1)\) global charge \(Q\) that we use as a controlling parameter.
For example in \(d + 1\) dimensions, we find that the leading contribution is \(E \sim Q^{(d+2)/d}\) \emph{vs.} \(E \sim Q^{(d+1)/d}\) in relativistic systems.
We also see that the speed of propagation of small fluctuations scales like \(c_s \sim Q^{2/d}\) \emph{vs.} \(c_s \sim Q^0\) in the relativistic case.

\medskip

Our result for a Schrödinger-invariant system with a global $U(1)$ symmetry in $3+1$ space-time dimensions must be compared to the results of Son and Wingate for the unitary Fermi gas~\cite{Son:2005rv}. They perform a small-momentum expansion for a superfluid %
at fixed chemical potential $\mu$, resulting in an effective action with leading and \ac{nlo} contributions.
The main difference is that we use the global charge as a dimensionless controlling parameter and do not need to make assumptions on the momenta. 
Since in both cases, Schrödinger invariance is imposed, we find the same operators that appear in their formula (116). %

\medskip

There are a number of further directions to pursue in the study of the large-charge expansion of non-relativistic systems. To make contact to the case of the anyons in $2+1$ dimensions, it is necessary to study the gauging of the global $U(1)$ symmetry. 

To be able to find anomalous operator dimensions, on the other hand, the system must be studied in a harmonic potential in order to make use of the non-relativistic state-operator correspondence described in~\cite{Nishida:2007pj}.
We leave these problems for future study.

\subsection*{Acknowledgements}

We would like to thank Simeon Hellerman and Uwe-Jens Wiese for enlightening discussions and detailed comments on the draft. %
D.O. and S.R. gratefully acknowledge support from the Simons Center for Geometry and Physics, Stony Brook University at which some of the research for this paper was performed.
The work of S.R. is supported by the Swiss National Science Foundation under grant number \textsc{pp00p2\_157571/1}.
D.O. acknowledges partial support by the \textsc{nccr 51nf40--141869} ``The Mathematics of Physics'' (Swiss\textsc{map}).

\bigskip

\appendix
\section{Schrödinger symmetry}%
\label{sec:schroedinger-symm}

The Schödinger Lagrangian of Eq~\eqref{eq:action_Schrödinger_Free_ddim} is invariant under the scale transformation
\begin{equation}
(t,x_i)  \rightarrow (t',x_i') = (e^{2\tau}t,e^{\tau} x_i),
\label{eq:nonrela_dila_transfo}
\end{equation}
where $\tau$ is a real parameter. In addition, the Schrödinger Lagrangian is also symmetric under the non-relativistic special conformal transformation 
\begin{equation}
(t,x_i)  \rightarrow (t', x_i') = \left(\frac{t}{1 + \lambda t}, \frac{x_i}{1 + \lambda t}\right)
\label{eq:nonrela_SCT} 
\end{equation}
 with $\lambda$ a real parameter.  The algebra that contains the Galilean algebra with central extension plus scale and special conformal transformations is called Schrödinger algebra. %
In~\cite{Nishida:2007pj} the generators of the Schrödinger algebra are given for a general Hamiltonian $H$: 
\begin{equation}
  \label{eq:Schroedinger-generators}
  \begin{aligned}
 N        & = \int \dd[d]{x} \rho(x)                    &  & \text{Mass}            \\
 P_i      & = \int \dd[d]{x} j_i(x)                  &  & \text{Momenta}         \\
 J_{ij}   & = \int \dd[d]{x} (x_i j_j(x)-x_j j_i(x)) &  & \text{Angular momenta} \\
 K_{i}    & = \int \dd[d]{x} x_i \rho(x)                &  & \text{Galilean boosts} \\
 D        & = \int \dd[d]{x} x_i j_i      &  & \text{Dilatation}      \\ 
 C        & = \int \dd[d]{x} \frac{x^2}{2} \rho(x)      &  & \text{Special conformal transformation}.
\end{aligned}
\end{equation}
The action of the free Schrödinger field in $d$ dimensions,
\begin{equation}\label{eq:action_Schrödinger_Free_ddim}
   S = \int \dd[d]{x} \dd{t} \frac{i}{2}\left(\psi^{*} \partial_t \psi -\psi \partial_t \psi^{*} \right) - \frac{\hbar}{2m}\partial_i \psi^{*} \partial_i \psi, 
\end{equation}
is invariant under non-relativistic conformal transformations.

\bigskip

It is convenient to consider the infinitesimal form of the scale and special conformal transformations (the expression for finite transformations are collected in Table~\ref{tab:summary_transfo}):
\begin{center}
\begin{tabular}{lrrr}
  \toprule
  & \(\delta t\) & \(\delta x^i\)& \(\delta \psi\) \\
  \midrule
  scale & \(2t\) & \(x^i\) & \(- \frac{d}{2} \psi \) \\
  \ac{sct} & \(-t^2\) & \(-t x^i\) & \( \frac{1}{2 } \pqty{ t d  - i \frac{m}{\hbar} x^2 } \psi \) \\
  \bottomrule
\end{tabular}
\end{center}
It is immediate to verify that the derivatives with respect to time and space and the volume element transform as
\begin{center}
\begin{tabular}{lrrr}
  \toprule
  & \(\delta \del_t\) & \(\delta \del_i\)& \(\delta (\dd{t}\dd[d]{x})\) \\
  \midrule
  scale & \(-2 \del_t\) & \(- \del_i\) & \( (d+2) \dd{t}\dd[d]{x} \) \\
  \ac{sct} & \(2 t \del_t + x^i \del_i\) & \(t \del_i\) & \( - (d + 2) t \dd{t}\dd[d]{x}\) \\
  \bottomrule
\end{tabular}
\end{center}
It is also convenient to decompose the complex field \(\psi\) into its radial and angular part \(\psi = a e^{-i \theta}\) which transform separately and consider the operator \(U = \del_t \theta - \hbar/(2m) \del_i \theta \del_i \theta \):
\begin{center}
\begin{tabular}{lrrr}
  \toprule
  & \(\delta a\) & \(\delta \theta \) & \( \delta U\)\\
  \midrule
  scale & \(- \frac{d}{2}  a \) & \(0\) & \(- 2 U\)\\
  \ac{sct} & \( t \frac{d}{2} a \) & \( \frac{m}{2 \hbar } x^2 \) & \(2 t U \)\\
  \bottomrule
\end{tabular}
\end{center}

The terms allowed in the effective action have to be invariant under the Schrödinger group.
This means that they have to be Galilei scalars (all indices are to be contracted with \(\delta_{ij}\) because parity forbids \(\epsilon_{ijk}\)), and invariant under scale transformations and \ac{sct}.
If we assume that there are no dimensionful couplings, scale invariance is implied by dimensional analysis.
To see that, observe that the generic term has the schematic form
\begin{equation}
  \mathcal{O} = \hbar^\alpha m^\beta \del_t^\gamma \del_i^\delta \theta^\epsilon
\end{equation}
and must have the dimensions of a Lagrangian density
\begin{equation}
  [ \mathcal{O}] = L^{2 - d} M T^{-2} .
\end{equation}
This fixes the coefficients \(\gamma\) and \(\delta\) to obey the relation
\begin{equation}
  2 \gamma + \delta = d + 2 .
\end{equation}
Under scale transformation, \(\mathcal{O}\) transforms as
\begin{equation}
  \mathcal{O} \mapsto \mathcal{O} + \tau \pqty{-2 \gamma - \delta} \mathcal{O}
\end{equation}
 and this compensates the variation of the volume element \(\dd{t}\dd[d]{x} \mapsto \dd{t}\dd[d]{x} + \tau \pqty{d + 2} \dd{t}\dd[d]{x}\) precisely when \(2\gamma + \delta  = d + 2\).

\begin{table}
\centering
\begin{tabular}{lrrr}
  \toprule
  & galilean transformations & Dilatations & \acp{sct} \\
  \midrule
   $(\vec{x},t)$ & $(R\vec{x}+\vec{v}t+\vec{a},t+b) $ & $(e^{\tau} \vec{x}, e^{2\tau} t)$ & $ \left(\frac{x}{1+\lambda t}, \frac{t}{1+\lambda t} \right)$\\
   $\psi(\vec{x},t)$ & $e^{\frac{i}{\hbar}\left(  m \vec{v} R \vec{x} +\frac{1}{2} m \vec{v}^2 t\right)} \psi(\vec{x},t) $ & $e^{\left(- \frac{d}{2} \tau \right)}\psi(\vec{x},t) $  & $ (1+ \lambda t)^{\frac{d}{2}}e^{\left(\frac{-i}{2\hbar}\frac{mx^2\lambda}{1+\lambda t}\right)}\psi(\vec{x},t) $ \\
   $\partial_t$  & $\partial_t -\vec{v}R^{-1} \nabla$ & $e^{-2\tau}\partial_t $ & $(1+\lambda t)^2 \partial_t + \lambda (1+\lambda t) \vec{x} \cdot \nabla$ \\   
   $\nabla$ & $R^{-1} \nabla$ & $e^{-\tau} \nabla$ & $(1+\lambda t) \nabla$ \\
   $\dd[d]{x} \dd{t}$  & $\dd[d]{x} \dd{t}$ & $ e^{\tau(d+2)}\dd[d]{x} \dd{t}$ & $(1+\lambda t)^{-(d+2)}\dd[d]{x} \dd{t}$\\
   $\theta(\vec{x},t)$  & $\theta(\vec{x},t) - \frac{1}{\hbar}\left(  m \vec{v} R \vec{x} +\frac{1}{2} m  \vec{v}^2 t\right)$ & $\theta(\vec{x},t)$ & $\theta(\vec{x},t) +\frac{1}{2\hbar}\frac{mx^2\lambda}{1+\lambda t} $ \\
   $U$ &$ U$ & $e^{-2\tau}U$ & $(1+ \lambda t)^2 U$ \\
  \bottomrule
\end{tabular}
\caption{Summary of the transformations under the Schrödinger group in $d$ space dimensions.}
 \label{tab:summary_transfo}
\end{table}

\setstretch{1}

\printbibliography{}

\end{document}